\documentclass[%
prb
,twocolumn%
,superscriptaddress
,amsmath,amssymb,aps,nobibnotes,
nofootinbib
,preprintnumbers
]{revtex4-1}
\usepackage[dvipdfmx]{graphicx}
\usepackage{bm}
\usepackage{dcolumn}
\usepackage{color}
\usepackage{tabularx}
\def\urm#1{\scriptstyle{\text{\textrm{\textmd{\textup{#1}}}}}}

\newcolumntype{Y}{>{\centering\arraybackslash}X}
\begin{document}
\title{Functional-renormalization-group aided density-functional analysis \\
  for the correlation energy of the two-dimensional homogeneous electron gas}
\author{Takeru Yokota}
\email{tyokota@ruby.scphys.kyoto-u.ac.jp}
\affiliation{Department of Physics, Faculty of Science, Kyoto University, Kyoto 606-8502, Japan}
\author{Tomoya Naito}
\email{naito@cms.phys.s.u-tokyo.ac.jp}
\affiliation{Department of Physics, Graduate School of Science, The University of Tokyo, Tokyo 113-0033, Japan}
\affiliation{RIKEN Nishina Center, Wako 351-0198, Japan}
\date{\today}
\preprint{KUNS-2743, RIKEN-QHP-387, RIKEN-iTHEMS-Report-18}
%
\begin{abstract}
  The functional-renormalization-group aided density-functional theory (FRG-DFT) is applied to 
  the two-dimensional homogeneous electron gas (2DHEG).
  The correlation energy of the 2DHEG is derived as a function of the Wigner-Seitz radius $ r_{\urm{s}} $ 
  directly.
  We find that our correlation energy completely reproduces the exact behavior 
  at the high-density limit.
  For finite density, 
  the result of FRG-DFT shows good agreement with the Monte Carlo (MC) results in the high-density region, 
  although the discrepancy between FRG-DFT and MC results becomes larger as the system becomes more dilute.
  Our study is the first example in which the FRG-DFT is applied to more-than-one-dimensional models,
  and shows that the FRG-DFT is a feasible and promising method even for the analysis of realistic models for quantum many-body systems.
\end{abstract}
\maketitle
\section{Introduction}
\par
The density functional theory (DFT) \cite{hoh64,koh65} is one of the most widely used methods
for the analysis of quantum many-body systems including electron systems.
The DFT is particularly known as a powerful tool for the analysis of the ground states
thanks to the Kohn-Sham scheme.\cite{koh65}
In the DFT, the energy density functional (EDF) plays a key role,
since the accuracy of the DFT calculation depends only on the accuracy of the EDF.
The Hohenberg-Kohn (HK) theorem \cite{hoh64} guarantees the existence
of such EDFs.
The recipe to get EDFs from the microscopic Hamiltonians is, however, not provided by the HK theorem,
and the construction of such a scheme is still an open problem.\cite{per01}
\par
A remarkable finding for the construction of the EDF 
is that the effective-action formalism gives a microscopic definition of the EDF.\cite{fuk94}
The effective actions,
which give the physical solutions at their stationary points,
are the quantum counterparts of the classical actions and include quantum fluctuations.
The effective action for the density, 
which is also known as the two-particle point irreducible effective action,\cite{ver92,bra12}
is given by the functional Legendre transformation of
a generating functional with a source coupled to the local composite density operator
$ \hat{\rho} \left( \bm{x} \right) = \hat{\psi}^\dagger \left(\bm{x} \right) \, \hat{\psi} \left(\bm{x} \right) $
with respect to the source.
It was found that the effective action for the density field
is corresponding to the EDF,\cite{fuk94} and the correspondence gives a microscopic definition of the EDF.
\par
Here, let us call the functional renormalization group (FRG),\cite{weg73,wil74,pol84,wet93}
which is an exact formalism for the renormalization-group procedure.
In particular, the FRG has been developed 
as a powerful computational machinery
in the effective-action formalism.\cite{wet93}
The FRG in the effective-action formalism provides non-perturbative and systematic ways
to construct effective actions based on microscopic Hamiltonians
by use of the one-parameter functional differential equation, 
which represents the change of the effective action
when the quantum fluctuations are gradually taken from the ultraviolet scale to the infrared scale;
see Refs.~\onlinecite{ber02,paw07,gie12,met12} for reviews.
Since EDFs can be constructed from the effective actions, 
a scheme to derive EDFs microscopically
is expected to be obtained by borrowing the idea from the FRG.
\par
The attempt to derive EDFs employing the idea of the FRG, which we call the functional-renormalization-group aided density-functional theory (FRG-DFT),
was initiated by Polonyi, Sailer, and Schwenk.\cite{pol02,sch04,pol05}
They discussed the functional evolution equation describing the change of the effective action 
when the interaction is gradually turned on.
Such a procedure of gradual turning on of the interaction is reminiscent of the adiabatic connection.\cite{har74,lan75,gun76,lan77,har84}
In the FRG-DFT, however, the EDF is obtained from the
functional differential equation in principle, 
in contrast to the most studies with the adiabatic connection,
in which some ansatz on the form of the functional,
or on the exchange-correlation kernel,
is needed to derive the EDF.
Moreover, flexible and systematic approximation schemes inspired by the FRG are provided in the FRG-DFT.
Recently, some applications have been demonstrated,\cite{kem13,lia18,kem17a,yok18,yok18b,ram17}
in some of which the FRG-DFT was found to describe the properties of the systems successfully.
These applications are, however, limited to the cases of 
zero-dimensional systems with and without the dimension of time,\cite{kem13,lia18}
and one-dimensional systems with short-range interactions
such as a simplified interaction representing the nuclear force\cite{kem17a,yok18,yok18b}
and the contact interaction.\cite{ram17}
\par
The FRG-DFT is applicable to infinite systems as well as finite systems.
The FRG-DFT formalism for the infinite systems has recently been developed in Ref.~\onlinecite{yok18},
in which the flowing chemical potential was introduced in order to control the expectation value of the particle number.
The formalism successfully described the properties of 
the ground state\cite{yok18} and excited states\cite{yok18b} of the one-dimensional homogeneous matter composed of spinless nucleons:\cite{ale89}
The equation of state (EOS) was calculated and the resultant saturation energy, which is the minimum of the EOS with respect to the density, showed a good agreement with the Monte Carlo (MC) calculation\cite{ale89}
in Ref.~\onlinecite{yok18}.
In addition, the density-density spectral function was derived in the framework of the FRG-DFT for the first time
and was found to reproduce some notable features of the non-linear Tomonaga-Luttinger liquid 
in Ref.~\onlinecite{yok18b}.
These results suggest that the FRG-DFT is a promising method for the analysis of excited states as well as the ground state.
\par
These successful analyses for infinite systems motivate one to apply the FRG-DFT to higher-dimensional realistic matters. 
The extension to higher-dimensional systems is formally straightforward,
though the FRG-DFT has not been applied to the more-than-one dimensional either finite or infinite systems.
A reasonable next step is applications to two-dimensional models in which the numerical analysis
are expected to be easier than those for three-dimensional models.
\par
In the condensed-matter-physics community, the two-dimensional homogeneous electron gas (2DHEG) has been studied\cite{tan89,kwo93,Polini2001J.Phys.Condens.Matter13_3591,att02,Luo2004Phys.Rev.Lett.92_047402,Ambrosetti2009Phys.Rev.B80_125306,Nagy2009NewJ.Phys.11_063012,loo11,Motta2015J.Chem.Phys.143_164108} for decades.
In spite of its simplicity, this model has played an important role for the understanding of physics emerged by the correlations of electrons,
and it shows rich physical phenomena, such as the Wigner crystallization and 
the magnetic phase transition.\cite{Ceperley1978Phys.Rev.B18_3126,tan89}
In addition, the two-dimensional electron gas is realized in, for instance, 
the semiconductor hetero-structures\cite{ando82}
and the atomic-layer materials,\cite{Novoselov2004Science306_666,Balendhran2014Small11_640}
and some interesting phenomena such as the quantum Hall effect\cite{Chang1984Phys.Rev.Lett.53_997,Nagaosa2010Rev.Mod.Phys.82_1539} have been studied in the two-dimensional electron systems.
The 2DHEG at the zero-temperature is parameterized by only two parameters, 
the Wigner-Seitz radius $ r_{\urm{s}} $ and the spin polarization $ \zeta $.
In this system, benchmark results for the correlation energy are available for wide range of densities:
At the high-density limit, the $ r_{\urm{s}} $-dependence of the correlation energy is known
to be described by the Gell-Mann-Brueckner resummation exactly.\cite{raj77}
For the case of finite density, the correlation energies were calculated by the MC calculations\cite{tan89,kwo93,att02,dru_nee09} at some $ r_{\urm{s}} $ and $ \zeta $.
\par
In this paper, we apply the FRG-DFT to the spin-unpolarized 2DHEG.
The correlation energy is derived as a function of $ r_{\urm{s}} $ with the second-order vertex expansion scheme.
We find that the correlation energy derived by the FRG-DFT 
completely reproduces the exact behavior
at the high-density limit ($ r_{\urm{s}} \to 0 $).
For finite $ r_{\urm{s}} $, we compare our results with those of the MC calculations.
The FRG-DFT result is in good agreement with MC results in the high-density region,
although the discrepancy between FRG-DFT and MC results becomes larger as $ r_{\urm{s}} $ increases.
Our study shows that the FRG-DFT is feasible even for two-dimensional systems and systems with the long-range interaction.
These results show that the FRG-DFT is a promising method to analyze realistic quantum many-body systems,
although the improvement of approximation is still required for more accurate description of dilute systems.
\par
This paper is organized as follows:
In Sec.~\ref{sec:form}, we present our formalism.
We derive the exact flow equation for the 2DHEG
and give the analytic solution of the correlation energy in the second-order vertex approximation in Sec.~\ref{sec:flow}.
We also discuss that the correlation energy derived by the FRG-DFT reproduces the exact behavior at the high-density limit from the formal viewpoint in Sec.~\ref{sec:exac}.
In Sec.~\ref{sec:nume}, we show the details for the numerical analysis.
We compare the result with those from the MC simulations.
Section~\ref{sec:conc} is devoted to the conclusion.
\section{Formalism\label{sec:form}}
\par
In this section, we present our formalism
to calculate the correlation energy of the spin-unpolarized 2DHEG.
We employ the FRG-DFT formalism for the infinite systems developed in Ref.~\onlinecite{yok18}.
The FRG-DFT flow equation and its solution in the second-order vertex expansion are shown.
We also discuss that the solution naturally reproduces the exact correlation energy at the high-density limit.
Here, in this paper, we employ the Hartree atomic unit.
\subsection{FRG-DFT flow equation\label{sec:flow}}
\par
We consider the spin-unpolarized 2DHEG at the zero temperature.
This system is characterized by only one parameter $ r_{\urm{s}} $,
which is defined by the electron density $ n = 1 / \left( \pi r_{\urm{s}}^2 \right) $.
The Hamiltonian of this system reads
\begin{align}
  \hat{H}
  = &
      \hat{H}_{\urm{el}}+\hat{H}_{\urm{el-i}}+\hat{H}_{\urm{i}},
      \label{eq:hami}\\
  \hat{H}_{\urm{el}}
  = &
      \sum_{s}\int_{\bm{x}}
      \hat{\psi}_{s}^* \left(\bm{x} \right) \,
      \frac{-\Delta}{2}
      \hat{\psi}_{s} \left( \bm{x} \right) \notag \\
    & +
      \frac{1}{2}
      \sum_{s,s'}
      \int_{\bm{x},\bm{x}'}
      \hat{\psi}_{s}^* \left( \bm{x} \right) \,
      \hat{\psi}_{s'}^* \left( \bm{x}' \right)
      \notag
      \\
      &
      \qquad \qquad \qquad
      \times
      U \left( \bm{x}-\bm{x}' \right) \,
      \hat{\psi}_{s'} \left( \bm{x}' \right) \,
      \hat{\psi}_{s} \left( \bm{x} \right),
      \notag \\
  \hat{H}_{\urm{el-i}}
  = &
      - n_{\urm{i}}
      \sum_{s} \int_{\bm{x},\bm{x}'}
      \hat{\psi}_{s}^* \left( \bm{x}\right) \,
      U \left( \bm{x}-\bm{x}' \right)  \,
      \hat{\psi}_{s} \left( \bm{x} \right),
      \notag \\
  \hat{H}_{\urm{i}}
  = & 
      \frac{n_{\urm{i}}^2}{2}
      \int_{\bm{x}, \bm{x}'}
      U \left (\bm{x}-\bm{x}' \right),
      \notag
\end{align}
where $ \bm{x} $ is the two-dimensional spacial coordinate,
$ \int_{\bm{x}} $ is the shorthand of $ \int d\bm{x} $,
$ \hat{\psi}_s \left( \bm{x} \right)$ and $ \hat{\psi}_s^* \left( \bm{x} \right) $ are the electron field operators with spin index $ s $,
$ U \left( \bm{x}-\bm{x}' \right) $ is the Coulomb interaction,
\begin{equation*}
  U \left( \bm{x}-\bm{x}' \right)
  :=
  \frac{1}{\left| \bm{x}-\bm{x}'\right|},
\end{equation*}
and $ n_{\urm{i}} = n $ is the density of the background opposite charge,
which has been introduced to neutralize the system.
Here, 
$ \hat{H}_{\urm{el}} $ contains the kinetic term for electrons and the electron-electron interaction term,
$ \hat{H}_{\urm{el-i}} $ is the electron-background interaction term, and
$ \hat{H}_{\urm{i}} $ is the interaction term of the background.
\par
We employ the imaginary-time path integral formalism for finite temperature for convenience, 
although we focus on the zero-temperature case in this paper.
Following the procedure in Refs.~\onlinecite{pol02,sch04},
we introduce the renormalization group (RG) parameter $ \lambda \in \left[ 0,1 \right] $
and the following $ \lambda $-dependent action: 
\begin{align}
  S_\lambda \left[ \psi^*,\psi \right]
  = &
      S_{\urm{el}, \lambda} \left[ \psi^*,\psi \right]
      + S_{\urm{el-i}, \lambda} \left[ \psi^*,\psi \right]
      + S_{\urm{i}, \lambda},
      \label{eq:acti} \\
  S_{\urm{el}, \lambda} \left[ \psi^*,\psi \right]
  = &
      \sum_{s}\int_{X}
      \psi_{s}^* \left( X_\epsilon \right) \,
      \left(
      \partial_\tau-\frac{1}{2}\Delta
      \right) \,
      \psi_{s} \left( X\right)
      \notag \\
    & +
      \frac{1}{2}
      \sum_{s,s'}\int_{X,X'}
      U_{\urm{2b}, \lambda} \left( X,X' \right)
      \notag \\
    & \times
      \psi_{s}^* \left( X_\epsilon \right) \,
      \psi_{s'}^* \left( X'_\epsilon \right) \,
      \psi_{s'} \left( X' \right) \,
      \psi_{s} \left( X \right),
      \notag \\
  S_{\urm{el-i}, \lambda} \left[ \psi^*,\psi \right]
  = &
      - n_{\urm{i}}
      \sum_{s}
      \int_{X,X'}
      U_{\urm{2b}, \lambda} \left( X,X' \right)
      \notag
      \\
      &\qquad\qquad\qquad
      \times
      \psi_{s}^* \left( X_\epsilon\right) \,
      \psi_{s} \left( X \right),
      \notag \\
  S_{\urm{i}, \lambda}
  = & 
      \frac{n_{\urm{i}}^2}{2}
      \int_{X,X'}
      U_{\urm{2b}, \lambda} \left (X,X' \right),
      \notag
\end{align}
where $ X = \left( \tau, \bm{x} \right) $ is the vector of the imaginary time $ \tau $ and $ \bm{x} $,
$ \int_X $ is the shorthand of $ \int_{0}^{\beta} d\tau \int d\bm{x} $ with the inverse temperature $ \beta $,
$ \psi_s \left( X \right)$ and $ \psi_s^* \left( X \right) $ are the electron fields with spin index $ s $,
and 
$ U_{\urm{2b}, \lambda} \left( X,X' \right) $ 
is defined as
\begin{equation*}
  U_{\urm{2b}, \lambda} \left( X, X' \right)
  :=
  \lambda \delta \left( \tau - \tau' \right) \, U \left( \bm{x} - \bm{x}' \right).
\end{equation*}
Here, $ \epsilon $ is a positive infinitesimal which appears when
constructing the path integral formalism based on the normal-ordered Hamiltonian,\cite{alt10}
$ X_\epsilon $ is $ X_\epsilon = \left( \tau + \epsilon, \bm{x} \right) $,
and $ S_{\lambda=0} $ is the action for the non-interacting Fermion gas, 
while $ S_{\lambda=1} $ is for the 2DHEG with the actual coupling strength, which is corresponding to
the Hamiltonian Eq.~\eqref{eq:hami}.
\par
Here, we introduce the the effective action for the density field
$ \rho_\psi \left( X \right) = \sum_{s} \psi^*_{s} \left( X_\epsilon \right) \, \psi_{s} \left( X \right) $,
which can be related to the EDF.\cite{fuk94}
So as to define the effective action for the density, we introduce the generating functional for
the density correlation functions:
\begin{equation*}
  Z_\lambda \left[ J \right]
  =
  \int
  \mathcal{D}\psi^* \, \mathcal{D}\psi \,
  \exp
  \left(
    \int_X
    J\left( X \right) \,
    \rho_\psi \left( X \right)
    -
    S_\lambda \left[ \psi^*,\psi \right]
  \right).
\end{equation*}
The generating functional for the connected density correlation functions
is given by $ W_\lambda \left[ J \right] = \ln Z_\lambda \left[ J \right] $,
which gives the $ n $-point correlation function in the presence of the external field $ J $ as follows:
\begin{equation*}
  G_\lambda^{(n)}
  \left[ J \right]
  \left( X_1, \ldots, X_n \right)
  :=
  \frac{\delta^{n} W_\lambda \left[ J \right]}
  {\delta J \left( X_1 \right) \, \cdots \, \delta J \left( X_n \right)}.
\end{equation*}
Then, the effective action for the density field is defined by the Legendre transformation:
\begin{equation}
  \label{eq:gamm}
  \Gamma_\lambda \left[ \rho \right]
  :=
  \int_X J_{\urm{sup}, \lambda}
  \left[ \rho \right] \left( X \right) \, \rho \left( X \right)
  -
  W_\lambda \left[ J_{\urm{sup},\lambda} \left[\rho \right] \right],
\end{equation}
where $ J_{\urm{sup}, \lambda} \left[ \rho \right] \left(X\right)$ is the external field satisfying
\begin{equation}
  \label{eq:jsup}
  \left.
    \frac{\delta W_\lambda \left[ J \right]}
    {\delta J \left( X \right)}
  \right|_{J=J_{\urm{sup},\lambda} \left[ \rho \right]}
  =
  \rho \left( X \right).
\end{equation}
The relation between the EDF $ E_\lambda \left[ \rho \right] $ and $ \Gamma_\lambda \left[ \rho \right] $ is given by\cite{fuk94}
\begin{equation}
  E_{\lambda}
  \left[ \rho \right]
  =
  \lim_{\beta \rightarrow \infty}
  \frac{\Gamma_{\lambda} \left[ \rho \right]}{\beta}.
  \label{eq:egam}
\end{equation}
This follows from the fact that 
the ground-state density $ \rho_{\urm{gs}, \lambda} \left( X \right) $ is given by the variational equation
\begin{equation}
  \label{eq:qeom}
  \left.
    \frac{\delta{\Gamma_\lambda} \left[ \rho \right]}
    {\delta \rho \left( X \right)}
  \right|_{\rho = \rho_{\urm{gs}, \lambda} \left( X \right)}
  = J_{\urm{sup},\lambda} \left[ \rho_{\urm{gs}, \lambda} \right] \left( X \right)
  = \mu_\lambda,
\end{equation}
and $ \lim_{\beta \rightarrow \infty} \Gamma_{\lambda} \left[ \rho_{\urm{gs}, \lambda} \right]/\beta $
can be identified with the ground-state energy $ E_{\urm{gs}, \lambda} $.
Here, we have introduced the $ \lambda $-dependent chemical potential $ \mu_\lambda $
satisfying $ \mu_{\lambda = 0} = \pi n =  1/r_{\urm{s}}^2 $
so that the electron density is fixed to $ n $ for all $ \lambda \in \left[ 0,1 \right] $.\cite{yok18}
The same chemical potential $ \mu_\lambda $ has been introduced
for the spin-up and spin-down electrons to make the particle number of electrons with each spin the same since the spin-unpolarized 2DHEG is considered.
\par
The change of $ \Gamma_\lambda\left[ \rho \right] $ with respect to $ \lambda $ is described by the following renormalization-group flow equation:\cite{sch04,kem17a}
\begin{widetext}
  \begin{align}
    \partial_\lambda \Gamma_\lambda \left[ \rho \right]
    = & 
        \frac{1}{2}
        \int_{X,X'}
        \partial_\lambda U_{\urm{2b}, \lambda} \left( X, X' \right) \, 
        \left(\rho \left( X \right) - n_{\urm{i}} \right)
        \left(\rho \left( X' \right) - n_{\urm{i}} \right) \notag \\
      & +
        \frac{1}{2}
        \int_{X,X'}
        \partial_\lambda U_{\urm{2b}, \lambda} \left( X, X' \right) \,
        \left(
        \left(
        \frac{\delta^2 \Gamma_{\lambda} \left[ \rho \right]}{\delta \rho \left( X_{\epsilon'} \right) \, \delta \rho \left( X' \right)}
        \right)^{-1}
        -
        \rho \left( X \right) \, \delta^{(2)} \left( \bm{x}-\bm{x}' \right)
        \right).
        \label{eq:flow}
  \end{align}
\end{widetext}
The first and second terms in the right-hand side correspond to the Hartree 
and exchange-correlation terms, respectively.
The positive infinitesimal $ \epsilon' $ satisfying $ \epsilon' > \epsilon $ 
has been introduced to avoid $ \tau = \tau' $ for 
$ \left(\delta^2 \Gamma_{\lambda} \left[ \rho \right]/\delta \rho \left( X \right) \, \delta \rho \left( X' \right)\right)^{-1} $, 
where its definition has uncertainty since $ \hat{\rho} $ also includes the infinitesimal $ \epsilon $ in its definition:
We have
\begin{align}
  & \left(
    \frac{\delta^2 \Gamma_{\lambda} \left[ \rho \right]}{\delta \rho \left( X \right) \, \delta \rho \left( X' \right)}
    \right)^{-1}        
    =
    \left.
    \frac{\delta^2 W_\lambda \left[ J \right]}
    {\delta J \left( X \right) \, \delta J \left( X' \right)}
    \right|_{J = J_{\urm{sup}, \lambda} \left[ \rho \right]}
    \notag \\
  & =
    \left\langle
    \hat{\rho} \left( X \right) \, \hat{\rho} \left( X' \right)
    \right\rangle_{\rho}
    -
    \rho \left( X \right) \, \rho \left( X' \right), \label{eq:gamw}
\end{align}
by use of
\begin{equation*}
  \frac{\delta^2 \Gamma_{\lambda} \left[ \rho \right]}
  {\delta \rho \left( X \right) \, \delta \rho \left( X' \right)}
  =
  \frac{\delta J_{\urm{sup}, \lambda} \left[\rho \right] \left( X \right)}{\delta \rho \left( X' \right)},
\end{equation*}
and
\begin{equation*}
  \left.
    \frac{\delta^2 W_\lambda \left[ J \right]}{\delta J \left( X \right) \, \delta J \left( X' \right)}
  \right|_{J=J_{\urm{sup}, \lambda} \left[ \rho \right]}
  =
  \left.
    \frac{\delta \rho \left( X \right)}{\delta J \left( X' \right)}
  \right|_{J=J_{\urm{sup}, \lambda} \left[ \rho \right]},
\end{equation*}
which are obtained from Eqs.~\eqref{eq:gamm} and \eqref{eq:jsup}, respectively.
Here, the average of an operator $ \hat{O} $
\begin{equation*}
  \left\langle
    \hat{O}
  \right\rangle_{\rho}
  :=
  \frac{
    \int \mathcal{D}\psi^* \, \mathcal{D}\psi \,
    \hat{O} 
    e^{
      \int_X
      J_{\urm{sup}, \lambda} \left[ \rho \right]\left( X \right) \,
      \hat{\rho}\left( X \right)
      -
      S_\lambda \left[ \psi^*,\psi \right]}}
  {Z \left[J_{\urm{sup},\lambda} \left[ \rho \right] \right]},
\end{equation*}
gives the time-ordered average.
The definition of 
$ \left(
    \delta^2 \Gamma_{\lambda} \left[ \rho \right] / \delta \rho \left( X \right) \, \delta \rho \left( X' \right) 
  \right)^{-1} $ at $\tau = \tau'$
is uncertain because the time ordering of
$ \psi_{s} \left( X \right) $ and $ \psi^*_{s'} \left( X_\epsilon' \right) $ in 
$ \left\langle \hat{\rho} \left( X \right) \, \hat{\rho} \left( X' \right)\right\rangle_{\rho}
=
\sum_{s,s'}\left\langle
  \psi^*_{s} \left( X_\epsilon \right) \, \psi_{s} \left( X \right) \,
  \psi^*_{s'} \left( X'_\epsilon \right) \, \psi_{s'} \left( X' \right) \,
\right\rangle_{\rho} $
depends on which limit $ \epsilon \to 0 $ or $ \tau \to \tau' $ is taken first,
i.e.,
\begin{equation*}
  \lim_{\tau \to \tau'}
  \lim_{\epsilon \to 0}
  \left\langle
    \hat{\rho} \left( X \right) \, \hat{\rho} \left( X' \right)
  \right\rangle_{\rho}
  \neq
  \lim_{\epsilon \to 0}
  \lim_{\tau \to \tau'}
  \left\langle
    \hat{\rho} \left( X \right) \, \hat{\rho} \left( X' \right)
  \right\rangle_{\rho}.
\end{equation*}
In order to make $
\left(
  \delta^2 \Gamma_{\lambda} \left[ \rho \right] / \delta \rho \left( X \right) \, \delta \rho \left( X' \right)
\right)^{-1} $
satisfy the ordering corresponding to the density-density correlation function in the flow equation,
we have introduced $ \epsilon' $.
Such a prescription yields the 
the term with the delta function in the last term; see Ref.~\onlinecite{yok18} for a detail.
\par
In principle, the effective action for the 2DHEG $ \Gamma_{\lambda=1} \left[ \rho \right] $
is obtained by solving Eq.~\eqref{eq:flow} starting from 
the effective action for the non-interacting system 
$ \Gamma_{\lambda=0} \left[ \rho \right] $.
Equation \eqref{eq:flow} is, however, a functional differential equation,
which is hard to be solved computationally and needs to be reduced to some numerically solvable equations for the practical use.
One of the schemes to realize such a reduction is the vertex expansion, 
in which the following Taylor series expansion
of $ \Gamma_\lambda \left[ \rho \right] $ around $ \rho = \rho_{\urm{gs}, \lambda} $
is employed:
\begin{widetext}
  \begin{equation*}
    \Gamma_{\lambda} \left[ \rho \right]
    = 
    \Gamma_\lambda \left[ \rho_{\urm{gs}, \lambda} \right]
    +
    \mu_\lambda 
    +
    \sum_{n=2}^{\infty}
    \frac{1}{n!}
    \int_{X_{1}} \cdots \int_{X_{n}}
    \frac{\delta^{n} \Gamma_\lambda[\rho_{\urm{gs}, \lambda}]}{\delta \rho(X_1)\cdots\delta \rho(X_n)} \,
    \left( \rho \left( X_1 \right) - \rho_{\urm{gs}, \lambda} \left( X_1 \right) \right)
    \cdots
    \left( \rho \left( X_n \right) - \rho_{\urm{gs}, \lambda} \left( X_n \right) \right).
  \end{equation*}
  By applying this expansion,
  Eq.~\eqref{eq:flow} is converted to a series of flow equations.
  Since the flow equation for $ \Gamma_\lambda^{(n)} $, which is the $ n $-th derivative of $ \Gamma_\lambda \left[ \rho \right] $ with respect to $ \rho $, depends on $ \Gamma_\lambda^{(m\leq n+2)} $, 
  these flow equations for $ \Gamma_{\lambda}^{(n)} $ form an infinite series of coupled differential equations.
  Therefore, a truncation for this series at some order is needed in practice.
  \par
  The ground-state energy and density,
  $ E_{\urm{gs}, \lambda} $ and $ \rho_{\urm{gs},\lambda} $,
  are related to $ \Gamma_\lambda \left[\rho \right] $
  via Eqs.~\eqref{eq:egam} and \eqref{eq:qeom},
  and the relations between the density correlation functions
  $ \left\lbrace
    G_\lambda^{(n)} \left( X_1, \ldots, X_n \right)
  \right\rbrace_{n=2}^{\infty}
  :=
  \left\lbrace
    G_\lambda^{(n)} \left[ J=\mu_\lambda \right] \left( X_1, \ldots, X_n \right)
  \right\rbrace_{n=2}^{\infty}$
  and the derivatives of $ \Gamma_\lambda \left[ \rho \right] $ 
  are derived from the derivatives of Eq.~\eqref{eq:gamw}
  with respect to $ \rho $.
  With these relations, the expansion up to the second order gives the flow equations for $ E_{\urm{gs}, \lambda} $, $ \rho_{\urm{gs}, \lambda} $, and the two-point density-correlation function $ G_{\lambda}^{(2)} $.
  These flow equations read
  \begin{align}
    \partial_\lambda E_{\urm{gs}, \lambda}
    = & 
        \lim_{\beta \to \infty}
        \frac{1}{\beta}
        \left[
        \int_X
        \mu_\lambda \, 
        \partial_\lambda \rho_{\urm{gs}, \lambda} \left( X \right)
        +
        \frac{1}{2}
        \int_{X,X'}
        \partial_\lambda U_{\urm{2b}, \lambda} \left( X, X' \right) \,
        \left(\rho_{\urm{gs}, \lambda} \left( X \right) - n_{\urm{i}} \right)
        \left(\rho_{\urm{gs}, \lambda} \left( X' \right) - n_{\urm{i}} \right)
        \right.
        \notag \\
      & + 
        \left.
        \frac{1}{2}
        \int_{X,X'}
        \partial_\lambda U_{\urm{2b}, \lambda} \left( X, X' \right) \,
        \left(
        G_{\lambda}^{(2)} \left( X_{\epsilon'}, X'\right)
        -
        \rho_{\urm{gs}, \lambda} \left( X' \right) \,
        \delta^{(2)} \left(\bm{x}-\bm{x}' \right)
        \right)
        \right],
        \label{eq:flo1} \\
    \partial_\lambda \rho_{\urm{gs}, \lambda} \left( X \right)
    = &
        \int_{X'}
        G_{\lambda}^{(2)} \left( X, X' \right) \,
        \left(
        \partial_\lambda \mu_\lambda
        -
        \int_{X''}
        \partial_\lambda U_{\urm{2b}, \lambda} \left( X', X'' \right) \,
        \left(\rho_{\urm{gs}, \lambda} \left( X'' \right) - n_{\urm{i}} \right)
        \right)
        \notag \\
      & -
        \frac{1}{2}
        \int_{X',X''}
        \partial_\lambda
        U_{\urm{2b}, \lambda} \left( X', X'' \right) \,
        \left(
        G_{\lambda}^{(3)} \left(X, X'_{\epsilon'}, X'' \right)
        -
        G_{\lambda}^{(2)} \left( X, X' \right) \,
        \delta^{(2)} \left(\bm{x}'-\bm{x}'' \right)
        \right),
        \label{eq:flo2} \\
    \partial_\lambda
    G_{\lambda}^{(2)} \left( X_1, X_2 \right)
    = & 
        \int_{X'}
        G_{\lambda}^{(3)} \left( X_1, X_2, X' \right) \,
        \left(
        \partial_\lambda \mu_\lambda
        -
        \int_{X''}
        \partial_\lambda U_{\urm{2b}, \lambda} \left( X', X'' \right) \,
        \left( \rho_{\urm{gs}, \lambda} \left( X'' \right) - n_{\urm{i}} \right)
        \right)
        \notag \\
      & - 
        \int_{X',X''}
        \partial_\lambda
        U_{\urm{2b}, \lambda} \left( X', X'' \right)
        \notag \\
      & \times
        \left[
        G_{\lambda}^{(2)} \left( X_1, X' \right) \,
        G_{\lambda}^{(2)} \left( X_2, X'' \right) 
        +
        \frac{1}{2}
        \left(
        G_{\lambda}^{(4)} \left( X_1, X_2, X', X''_{\epsilon'} \right)
        -
        G_{\lambda}^{(3)} \left( X_1, X_2, X'' \right) \,
        \delta^{(2)} \left( \bm{x}'-\bm{x}'' \right)
        \right)
        \right].
        \label{eq:flo3}
  \end{align}
\end{widetext}
\par
We have mentioned that 
$ \mu_\lambda $ is chosen so that the density of the system is fixed to $ n $ during the flow,
i.e., $ \rho_{\urm{gs}, \lambda=0} \left( X \right) = n $
and $ \partial_\lambda \rho_{\urm{gs}, \lambda} \left( X \right) = 0 $ for any $ \lambda $.
Here, we present how to choose such $ \mu_\lambda $
and derive the flow equations for the ground-state energy per particle and the two-point density-correlation function
with the choice of $ \mu_\lambda $, as shown in Ref.~\onlinecite{yok18}.
Since we consider the homogeneous system,
the momentum representation as Ref.~\onlinecite{yok18} is convenient.
From Eq.~\eqref{eq:flo2}, one finds that
$ \rho_{\urm{gs}, \lambda} \left( X \right) = n $, i.e.
$ \partial_\lambda \rho_{\urm{gs}, \lambda} \left( X \right) = 0 $,
is realized by the following choice of $ \mu_\lambda $:
\begin{align*}
  & \partial_\lambda \mu_\lambda \\
  & = 
    \frac{1}{2 \tilde{G}_{\lambda}^{(2)} \left( 0 \right)}
    \int_{\bm{p}}
    \tilde{U} \left( \bm{p} \right)
    \left(
    \int_\omega
    e^{i \omega \epsilon'}
    \tilde{G}_{\lambda}^{(3)} \left( P, -P\right)
    -
    \tilde{G}_{\lambda}^{(2)} \left( 0 \right)
    \right),
\end{align*}
where we have introduced 
$ \int_{\bm{p}} := \int d{\bm p} / \left( 2 \pi \right)^2 $, 
$ \int_{\omega} := \int d \omega / \left( 2 \pi \right) $, 
$ \int_{P} := \int_{\bm{p}} \int_{\omega} $, and
the Fourier transformations of $ U \left( \bm{x} \right) $
and
$ G^{(n)}_\lambda \left( X_1, \ldots, X_n \right) $:
\begin{align*}
  \tilde{U} \left( \bm{p} \right)
  & :=
    \int_{\bm{x}}
    U \left( \bm{x} \right) \,
    e^{-i \bm{p} \cdot \bm{x}}
    =
    \frac{2 \pi}{\left| \bm{p} \right|}, \\
  \left( 2 \pi \right)^3
  & \delta^{(3)} \left( P_1 + \cdots + P_n \right) \,
    \tilde{G}_\lambda^{(n)} \left( P_1, \ldots, P_{n-1} \right) \\
  & :=
    \int_{X_1, \ldots, X_n}
    e^{-i
    \left(
    P_1 \cdot X_1 + \cdots + P_n \cdot X_n \right)}
    G_\lambda^{(n)} \left( X_1, \ldots, X_{n} \right).
\end{align*}
Here, we should note that $ \tilde{G}_\lambda^{(2)} \left( 0 \right) = \lim_{\bm{p} \to \bm{0}} \tilde{G}^{(2)}_\lambda \left( 0, \bm{p} \right) $ is 
interpreted as the $ p $ limit of $ \tilde{G}_\lambda^{(2)} \left( P \right)$,
which is the static particle-density susceptibility and generally nonzero.\cite{for75,kun91,fuj04,yok18}
Then, from Eqs.~\eqref{eq:flo1} and \eqref{eq:flo3},
the flow equations for the ground-state energy per particle $ E_{\urm{gs}, \lambda} / N $ and the two-point density-correlation function $ \tilde{G}_{\lambda}^{(2)} $ 
in terms of the momentum representation read 
\begin{align}
  \partial_\lambda \frac{{E}_{\urm{gs}, \lambda}}{N}
  & = 
    \frac{1}{2n}
    \int_{\bm{p}}
    \tilde{U} \left(\bm{p} \right) \,
    \left(
    \int_\omega
    e^{i\omega \epsilon'}
    \tilde{G}^{(2)}_{\lambda} \left( P \right)
    -
    n
    \right),
    \label{eq:fene} \\
  \partial_\lambda \tilde{G}_{\lambda}^{(2)} \left(P \right)
  & = 
    -
    \tilde{U} \left( \bm{p} \right) \,
    \left[
    \tilde{G}_{\lambda}^{(2)} \left( P \right)
    \right]^2
    + C_\lambda \left( P \right),
    \label{eq:fg2t}
\end{align}	
respectively. 
The first and second terms in the right-hand side of Eq.~\eqref{eq:fg2t}
come from the direct and exchange-correlation terms, respectively.
These terms are resummed by solving Eq.~\eqref{eq:fg2t}.
Here, $ N = n \int d \bm{x} $ is the number of electrons and
\begin{align}
  C_\lambda \left( P \right)
  := &
       \frac{1}{2}
       \int_{P'}
       e^{i \omega' \epsilon'}
       \tilde{U} \left( \bm{p}' \right)
       \frac{
       \tilde{G}_{\lambda}^{(3)} \left( P', -P' \right) \,
       \tilde{G}_{\lambda}^{(3)} \left( P, -P \right)}
       {\tilde{G}_{\lambda}^{(2)} \left( 0 \right)} \notag \\
     & -
       \frac{1}{2}
       \int_{P'}
       e^{i \omega' \epsilon'}
       \tilde{U} \left( \bm{p}' \right) \,
       \tilde{G}^{(4)}_{\lambda} \left( P', -P', P \right).
       \label{eq:C_lambda}
\end{align}
We note that the Hartree term is correctly canceled out with the positive background due to $ S_{\urm{i}, \lambda} $.\cite{mah00}
The ground-state energy per particle is obtained by integrating Eq.~\eqref{eq:fene} with respect to $ \lambda $:
\begin{widetext}
  \begin{equation}
    \frac{{E}_{\urm{gs}, \lambda = 1}}{N}
    =
    \frac{{E}_{\urm{gs}, \lambda = 0}}{N}
    +
    \frac{1}{2n}
    \int_{\bm{p}}
    \tilde{U} \left(\bm{p} \right) \,
    \left(
      \int_\omega
      e^{i \omega \epsilon'}
      \tilde{G}^{(2)}_{\lambda = 0} \left( P \right)
      -
      n
    \right)
    +
    \frac{1}{2n}
    \int_{0}^{1} d \lambda 
    \int_{P}
    e^{i \omega \epsilon'}
    \tilde{U} \left(\bm{p} \right) \,
    \left(
      \tilde{G}^{(2)}_{\lambda} \left( P \right)
      -
      \tilde{G}^{(2)}_{\lambda = 0} \left( P \right)
    \right).
    \label{eq:ediv}
  \end{equation}
  In the right-hand side,
  the first term is the energy for the free system,
  and identical with the kinetic energy given by $ 1 / \left( 2r_{\urm{s}}^2 \right) $.
  The second term is the exchange energy given by
  $ -4 \sqrt{2} / \left( 3 \pi r_{\urm{s}} \right) $.\cite{Slater1951Phys.Rev.81_385}
  The third term corresponds to the correlation energy:
  \begin{equation}
    \frac{E_{\urm{corr}}}{N}
    :=
    \frac{1}{2n}
    \int_{0}^{1} d\lambda
    \int_{P}
    e^{i \omega \epsilon'}
    \tilde{U} \left( \bm{p} \right) \,
    \left(
      \tilde{G}^{(2)}_{\lambda} \left( P \right)
      -
      \tilde{G}^{(2)}_{\lambda = 0} \left( P \right)
    \right).
    \label{eq:corr}
  \end{equation}
Note that the Hartree term does not appear in Eq.~\eqref{eq:ediv}
since the background opposite charge has cancelled it out.
  \par
  Here, $ \tilde{G}^{(2)}_{\lambda} \left( P \right) $ appearing in the right-hand side of Eq.~\eqref{eq:corr}
  is obtained by solving Eq.~\eqref{eq:fg2t}.
  However, 
  some approximation on $ C_\lambda \left( P \right) $ in the right-hand side of Eq.~\eqref{eq:fg2t}
  is needed
  since the flows of $ G_{\lambda}^{(3)} $ and $ G_{\lambda}^{(4)} $,
  which appear in the definition of $ C_\lambda \left( P \right) $ as Eq.~\eqref{eq:C_lambda},
  are not taken into account in the vertex expansion up to the second order.
  In this paper, we simply ignore the $ \lambda $-dependence of $ C_\lambda \left( P \right) $:
  $ C_\lambda \left( P \right) \approx C_{\lambda = 0} \left( P \right)$.
  Under this approximation, Eq.~\eqref{eq:fg2t} can be solved analytically.
  The solution reads
  \begin{equation}
    \tilde{G}_\lambda^{(2)} \left( P \right)
    =
    \frac{
      \tilde{G}_{\lambda = 0}^{(2)} \left( P \right)
      +
      \sqrt{C_{\lambda = 0} \left( P \right) / \tilde{U} \left(\bm{p} \right)}
      \tanh
      \left(
        \lambda \tilde{U} \left( \bm{p} \right) \,
        \sqrt{C_{\lambda = 0}\left( P \right) / \tilde{U} \left(\bm{p} \right)}
      \right)
    }{
      1
      +
      \sqrt{\tilde{U} \left( \bm{p} \right) / C_{\lambda = 0} \left( P \right)}
      \tilde{G}_{\lambda = 0}^{(2)} \left( P \right) \,
      \tanh
      \left(
        \lambda \tilde{U} \left( \bm{p} \right)
        \sqrt{C_{\lambda = 0} \left( P \right) / \tilde{U} \left(\bm{p} \right)}
      \right)}.
    \label{eq:gsol}
  \end{equation}
  Then, the integral with respect to $ \lambda $ in Eq.~\eqref{eq:corr} can be performed analytically.
  The resultant correlation energy is as follows:
  \begin{equation}
    \frac{E_{\urm{corr}}}{N}
    = 
    \frac{1}{2n}
    \int_{P}
    \left(
      \ln \left[
        \cosh \left(
          \sqrt{\tilde{U} \left( \bm{p}\right) \, C_{\lambda = 0} \left( P \right)}
        \right)
        +
        \sqrt{\frac{\tilde{U} \left( \bm{p} \right)}{C_{\lambda = 0} \left( P \right)}}
        \tilde{G}_{\lambda = 0}^{(2)} \left( P \right)
        \sinh \left(
          \sqrt{\tilde{U} \left( \bm{p} \right) \, C_{\lambda = 0} \left( P \right)}
        \right)	
      \right]
      -
      \tilde{U} \left( \bm{p} \right) \, \tilde{G}_{\lambda = 0}^{(2)} \left( P \right)
    \right).
    \label{eq:esol}
  \end{equation}
  \par
  We should specify $ \tilde{G}_{\lambda = 0}^{(2,3,4)} $ appearing in Eq.~\eqref{eq:esol}
  and $ C_{\lambda = 0} \left( P \right) $.
  The $ n $-point density-correlation function for the free case, $ \tilde{G}^{(n)}_{\lambda = 0} $, reads
  \begin{equation*}
    \tilde{G}^{(n)}_{\lambda = 0}
    \left( P_1, \ldots, P_{n-1} \right)
    = 
    -
    N_{\urm{s}}
    \sum_{\sigma \in S_{n-1}}
    \int_{P'}
    \prod_{k = 0}^{n-1}
    \tilde{G}_{\urm{F}, 0}^{(2)}
    \left(
      \sum_{i = 1}^{k}
      P_{\sigma(i)} + P'
    \right),
  \end{equation*}
  where $ N_{\urm{s}} = 2 $ is the spin degrees of freedom,
  $ S_{n-1} $ is the symmetric group of order $ n-1 $,
  and $ \sum_{i = 1}^{k=0} P_{\sigma(i)} = 0 $.
  Here, $ \tilde{G}_{\urm{F}, 0}^{(2)} \left( P \right) $ is the two-point propagator of free Fermions:
  $ \tilde{G}_{\urm{F}, 0}^{(2)} \left( P \right)
  = e^{i \omega \epsilon} / \left( i \omega - \xi \left( \bm{p} \right) \right) $,
  where $ \xi \left( \bm{p} \right) := \bm{p}^2/2 - \mu_{\lambda = 0} $.
  Using these expressions for the density-correlation functions, we have 
  \begin{align}
    \tilde{G}^{(2)}_{\lambda = 0}
    \left( P \right)
    = & 
        2 N_{\urm{s}}
        \int_{\bm{p}'}
        \theta \left( -\xi \left( \bm{p}' \right) \right) 
        \frac{
        \xi \left( \bm{p} + \bm{p}' \right) - \xi \left( \bm{p}' \right)}{
        \omega^2 + \left[ \xi \left( \bm{p}+ \bm{p}' \right) - \xi \left( \bm{p}' \right) \right]^2},
        \label{eq:g2fr} \\
    C_{\lambda = 0} \left( P \right)
    = & 
        2 N_{\urm{s}}
        \iint_{\bm{p}', \bm{p}''}
        U \left( \bm{p}' \right)
        \theta \left( -\xi \left(\bm{p}'' \right) \right) \,
        \left(
        \theta \left( -\xi \left( \bm{p} + \bm{p}' + \bm{p}'' \right) \right)
        -
        \theta \left( -\xi \left( \bm{p}' + \bm{p}'' \right) \right)
        \right)
        \notag \\
      & \times 
        \left[
        \frac{
        \left( \xi \left( \bm{p}'' + \bm{p} \right) - \xi \left( \bm{p}'' \right) \right)^2 - \omega^2}
        {\left( \omega^2 + \left( \xi \left(\bm{p}''+\bm{p} \right) - \xi \left( \bm{p}'' \right) \right)^2 \right)^2}
        -
        \frac{
        \left(
        \xi \left( \bm{p}'' + \bm{p} + \bm{p}' \right) - \xi \left( \bm{p}'' + \bm{p}'\right)
        \right)
        \left(
        \xi \left( \bm{p}'' + \bm{p} \right) - \xi \left( \bm{p}'' \right) \right)
        -\omega^2}
        {\left(
        \omega^2 + \left(
        \xi \left( \bm{p}'' + \bm{p} + \bm{p}' \right) - \xi \left( \bm{p}'' + \bm{p}' \right)
        \right)^2
        \right) \left(
        \omega^2 + \left(
        \xi \left( \bm{p}'' + \bm{p} \right) - \xi \left( \bm{p}'' \right) \right)^2
        \right)} \right],
        \label{eq:g4fr}
  \end{align}
  where $ \theta \left( x \right) $ is the Heaviside step function.
  We note that by use of the $ \tilde{G}^{(2)}_{\lambda = 0} \left( P \right) $ given in Eq.~\eqref{eq:g2fr},
  the exchange energy $ -4\sqrt{2} / \left(3 \pi r_{\urm{s}} \right) $
  is correctly obtained from the second term in the right-hand side of Eq.~\eqref{eq:ediv}.
\end{widetext}
%
\subsection{Reproduction of the exact correlation energy at the high-density limit
  \label{sec:exac}}
\par
We show how our solution Eq.~\eqref{eq:esol} behaves at the high-density limit ($ r_{\urm{s}} \to 0 $).
First, we discuss how
$ C_{\lambda = 0} \left( P \right) $ and $ \tilde{G}_{\lambda = 0}^{(2)} \left( P \right) $
depend on the Fermi momentum 
$ p_{\urm{F}} = \sqrt{2}/r_{\urm{s}} $ or the Wigner-Seitz radius $ r_{\urm{s}} $.
Since $ \mu_{\lambda = 0} = p_{\urm{F}}^2 / 2 $,
$ C_{\lambda = 0} $ and $ \tilde{G}_{\lambda = 0}^{(2)} $ have 
$ p_{\urm{F}} $-dependence through
$ \xi \left( \bm{p} \right) = \bm{p}^2/2 -p_{\urm{F}}^2/2$.
Here, $ C_{\lambda = 0} \left( P \right) $ and $ \tilde{G}_{\lambda = 0}^{(2)} \left( P \right) $
are redefined as 
$ C_{\lambda = 0} \left( \omega, \bm{p}; p_{\urm{F}} \right) $ 
and 
$ \tilde{G}_{\lambda = 0}^{(2)} \left( \omega, \bm{p}; p_{\urm{F}} \right) $, respectively,
to discuss the $ p_{\urm{F}} $-dependence of $ C_{\lambda = 0} $ and $ \tilde{G}_{\lambda = 0}^{(2)} $.
When the momentum and frequency are rescaled
as $ \overline{\omega} = \omega/p_{\urm{F}}^2 $
and $ \overline{\bm{p}}= \bm{p}/p_{\urm{F}} $,
$ C_{\lambda = 0} \left(\omega, \bm{p}; p_{\urm{F}} \right) $
and $ \tilde{G}_{\lambda = 0}^{(2)} \left(\omega, \bm{p}; p_{\urm{F}} \right) $
behave as follows:
\begin{align}
  C_{\lambda = 0} \left(\omega, \bm{p}; p_{\urm{F}} \right)
  & =
    p_{\urm{F}}^{-1}
    C_{\lambda = 0} \left(\overline{\omega}, \overline{\bm{p}}; 1 \right),
    \label{eq:cres} \\
  \tilde{G}_{\lambda = 0}^{(2)} \left(\omega, \bm{p}; p_{\urm{F}} \right)
  & = 
    \tilde{G}_{\lambda = 0}^{(2)} \left(\overline{\omega}, \overline{\bm{p}}; 1 \right).
    \label{eq:gres}
\end{align}
By use of these relations, Eq.~\eqref{eq:esol}
can be rewritten as
\begin{widetext}
  \begin{align}
    \frac{E_{\urm{corr}}}{N}
    = & 
        \frac{2 \pi}{r_{\urm{s}}^2}
        \int_{\overline{P}}
        \left(
        \ln \left[
        1
        +
        \sqrt{\frac{\tilde{U} \left(\overline{\bm{p}}\right)}
        {C_{\lambda = 0} \left(\overline{\omega}, \overline{\bm{p}}; 1\right)}}
        \tilde{G}_{\lambda = 0}^{(2)} \left(\overline{\omega}, \overline{\bm{p}}; 1 \right) \,
        \tanh
        \left(
        r_{\urm{s}}
        \sqrt{\frac{\tilde{U} \left(\overline{\bm{p}}
        \right)
        C_{\lambda = 0} \left(\overline{\omega}, \overline{\bm{p}}; 1 \right)}{2}}
        \right)
        \right]
        -
        \frac{r_{\urm{s}}}{\sqrt{2}}
        \tilde{U} \left(\overline{\bm{p}} \right) 
        \tilde{G}_{\lambda = 0}^{(2)} \left(\overline{\omega}, \overline{\bm{p}}; 1 \right)
        \right)
        \notag \\
            & + 
              \frac{2 \pi}{r_{\urm{s}}^2}
              \int_{\overline{P}}
              \ln \left[
              \cosh \left(
              r_{\urm{s}}
              \sqrt{\frac{\tilde{U} \left(\overline{\bm{p}} \right)
              C_{\lambda = 0} \left(\overline{\omega}, \overline{\bm{p}}; 1 \right)}{2}}
              \right)
              \right],
              \label{eq:eres}
  \end{align}
  where we have used $ p_{\urm{F}} = \sqrt{2} / r_{\urm{s}} $ and introduced 
  $
  \int_{\overline{P}} := \int d \overline{\omega}/ \left( 2 \pi \right) \int d\overline{\bm{p}} / \left( 2 \pi \right)^2 $.
  By expanding this equation with respect to $ r_{\urm{s}} $,
  we have
  \begin{align}
    \frac{E_{\urm{corr}}}{N}
    & = 
      \frac{2 \pi}{r_{\urm{s}}^2}
      \int_{\overline{P}}
      \left(
      \ln \left[
      1
      +
      r_{\urm{s}}
      \frac{\tilde{U} \left(\overline{\bm{p}} \right)}{\sqrt{2}}
      \tilde{G}_{\lambda = 0}^{(2)} \left(\overline{\omega}, \overline{\bm{p}}; 1 \right)
      \right]
      -
      \frac{r_{\urm{s}}}{\sqrt{2}}
      \tilde{U} \left(\overline{\bm{p}} \right) 
      \tilde{G}_{\lambda = 0}^{(2)} \left(\overline{\omega}, \overline{\bm{p}}; 1 \right)
      \right)
      +
      \pi
      \int_{\overline{P}}
      \frac{\tilde{U} \left(\overline{\bm{p}}\right)
      C_{\lambda = 0} \left(\overline{\omega}, \overline{\bm{p}}; 1 \right)}{2}
      +
      \mathcal{O} \left(r_{\urm{s}} \right)
      \notag \\
    & = 
      \frac{1}{2n}
      \int_{P}
      \left(
      \ln \left[
      1
      +
      \tilde{U} \left(\bm{p} \right)
      \tilde{G}_{\lambda = 0}^{(2)} \left(\omega, \bm{p}; p_{\urm{F}} \right)
      \right]
      -
      \tilde{U} \left(\bm{p} \right) 
      \tilde{G}_{\lambda = 0}^{(2)} \left(\omega, \bm{p}; p_{\urm{F}} \right)
      \right)
      +
      \frac{1}{4n}
      \int_{P}
      \tilde{U} \left(\bm{p} \right)
      C_{\lambda = 0} \left(\omega, \bm{p}; p_{\urm{F}} \right)
      +
      \mathcal{O} \left(r_{\urm{s}}\right).
      \label{eq:ecex}
  \end{align}
\end{widetext}
We note that $ \tilde{U} \left( \overline{\bm{p}} \right) $,
$ \tilde{G}_{\lambda = 0}^{(2)} \left( \overline{\omega}, \overline{\bm{p}}; 1 \right) $,
and
$ C_{\lambda = 0} \left(\overline{\omega}, \overline{\bm{p}}; 1 \right) $
do not depend on $ r_{\urm{s}} $.
\par
The first term of Eq.~\eqref{eq:ecex} is identical with the contributions from ring diagrams
and the second term is found to be the same as the contribution from the second-order exchange term.
Actually, by performing the frequency integral of the second term, we have
\begin{align*}
  \frac{1}{4n}
  & \int_{P}
    \tilde{U} \left( \bm{p} \right)
    C_{\lambda = 0} \left(\omega, \bm{p} ; p_{\urm{F}} \right) \\
  = & 
      \frac{2\pi^2 N_{\urm{s}}}{n}
      \iiint_{\bm{p}, \bm{p}', \bm{p}''}
      \frac{
      \theta \left(- \xi \left(\bm{p}'' \right) \right) \,
      \theta \left(- \xi \left(\bm{p}' \right) \right)}
      {\left| \bm{p} \right| \left| \bm{p}+\bm{p}'+\bm{p}'' \right|} \\
  & \times
    \frac{
    \left( 1 - \theta \left( - \xi \left( \bm{p} + \bm{p}' \right) \right) \right) \,
    \left( 1 - \theta \left( - \xi \left( \bm{p} + \bm{p}'' \right) \right) \right)}
    {\bm{p} \cdot \left(\bm{p} + \bm{p}' + \bm{p}'' \right)},
\end{align*}
which is identical with the expression of the second-order exchange contribution.\cite{raj77}
Therefore, the asymptotic form shown in Eq.~\eqref{eq:ecex} is 
the same as the expression given by the Gell-Man-Brueckner resummation,\cite{gel57,raj77}
and our correlation energy
naturally reproduces the exact behavior at the high-density limit:\cite{raj77,isi80,loo11}
\begin{align}
  \frac{E_{\urm{corr}}}{N}
  = &
      \ln 2 - 1 
      +
      \beta \left( 2 \right)
      -
      \frac{8}{\pi^2} \beta \left( 4 \right)
      \notag \\
    & -
      \sqrt{2}
      \left(
      \frac{10}{3 \pi} - 1
      \right)
      r_{\urm{s}} \ln r_{\urm{s}}
      +
      \mathcal{O} \left( r_{\urm{s}} \right)
      \notag \\
  = &
      - 0.192496 \ldots
      - 0.0863136 \ldots
      \times r_{\urm{s}}\ln r_{\urm{s}}
      +
      \mathcal{O} \left( r_{\urm{s}} \right),
      \label{eq:exact}
\end{align}
where $ \beta \left( x \right) $ is the Dirichlet beta function.
\section{Numerical results\label{sec:nume}}
\subsection{Details for the numerical calculation}
\par
In this subsection, we mention some details for the numerical calculation of Eqs.~\eqref{eq:esol}, \eqref{eq:g2fr}, and \eqref{eq:g4fr}.
\par
Equation \eqref{eq:g4fr} has a quadruple momentum integral.
In the case of the Coulomb interaction, however, this integral can be analytically reduced to a double integral,
which reduces the time for the numerical computation.
Further reduction of the computational time is possible by using the relation Eqs.~\eqref{eq:cres} and \eqref{eq:gres}:
Thanks to these relations,
$ C_{\lambda = 0} \left(\omega, \bm{p}; p_{\urm{F}} \right) $
and
$ \tilde{G}_{\lambda = 0}^{(2)} \left(\omega, \bm{p}; p_{\urm{F}} \right) $
are easily obtained for
various $ p_{\urm{F}} $, i.e.,~$ r_{\urm{s}} $,
once $ C_{\lambda = 0} \left(\overline{\omega}, \overline{\bm{p}}; 1 \right) $
and $ \tilde{G}_{\lambda = 0}^{(2)}\left( \overline{\omega}, \overline{\bm{p}}; 1 \right) $
are numerically obtained.
\par
Since the integrand in Eq.~\eqref{eq:esol} does not depend on the direction of the momentum,
the angular integration can be performed easily and the momentum integral is reduced to the integral with respect to $ \left| \bm{p} \right| $.
Moreover, the interval of the $ \omega $-integration can be restricted to $ \left[ 0, \infty \right) $
since the integrand in Eq.~\eqref{eq:esol} is an even function of $ \omega $.
For the numerical calculation, we change the variables for integral as
$ \theta_{\omega} = \left( 2 / \pi \right) \arctan \left( \omega^{\alpha_\omega} /s_\omega \right) $
and $ \theta_{p} = \left( 2 / \pi \right) \arctan \left( p^{\alpha_p} / s_p \right) $,
where $ \alpha_{\omega,p} $ and $ s_{\omega, p} $ are arbitral positive numbers.
Then, the interval of the numerical integrations are changed
from 
$ p , \, \omega \in \left[ 0, \infty \right) $
to $ \theta_{p} , \, \theta_{\omega} \in \left[ 0, 1 \right] $.
\subsection{Correlation energy}
\begin{figure}[!t]
  \centering
  \includegraphics[width=\columnwidth]{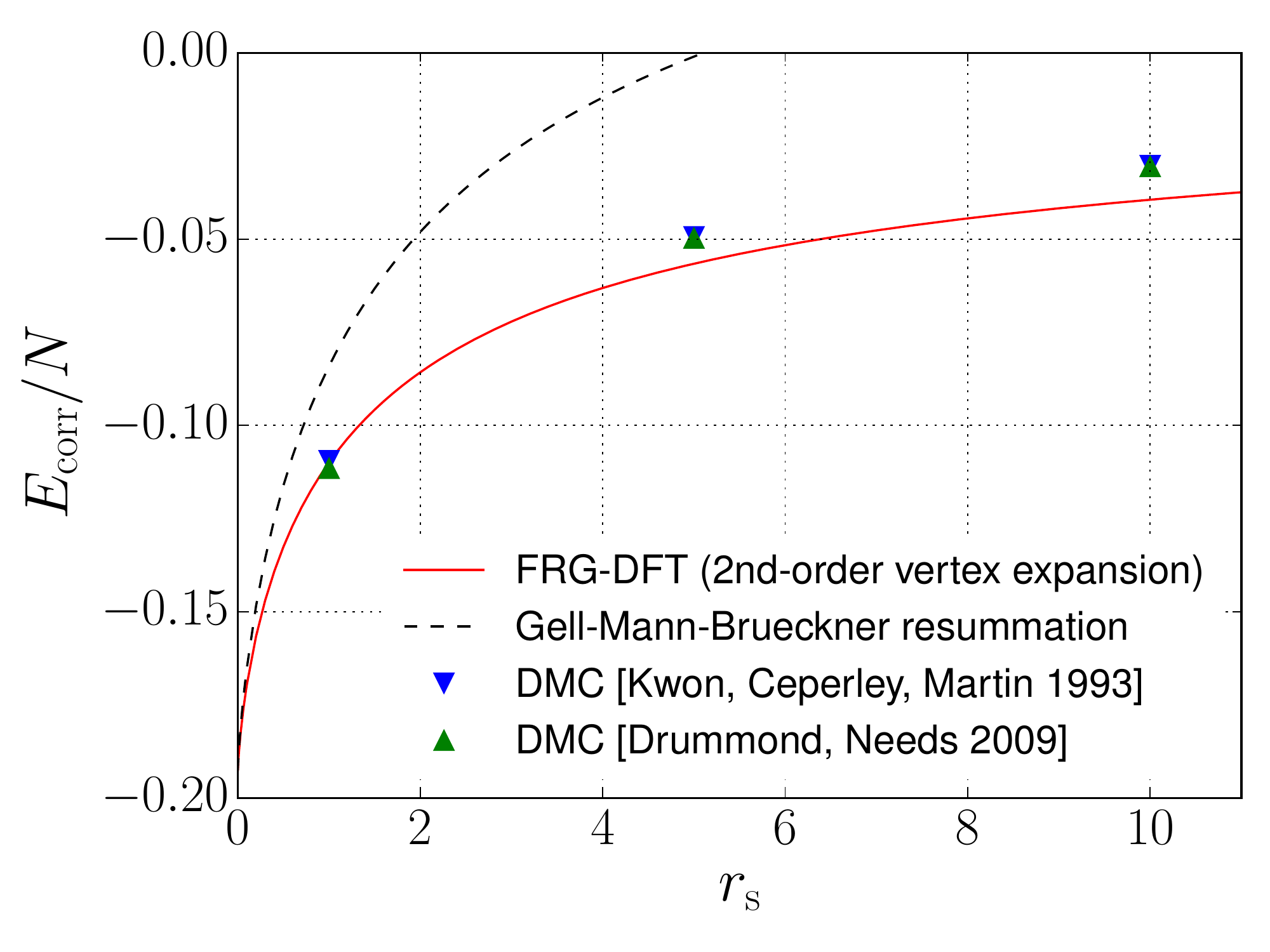}
  \caption{Correlation energy $ E_{\urm{corr}} / N $ of the 2DHEG derived by the FRG-DFT method (solid red line)
    shown as the function of the Wigner-Seitz radius $ r_{\urm{s}} $.
    For comparison, the results derived by the Gell-Mann-Brueckner resummation (black dashed line)
    and the Monte-Carlo calculations are also shown.
    The green inverted triangles and blue triangles are the energies derived by
    the extrapolations of the results of
    the DMC calculations to the infinite systems
    given by Kwon et al.\cite{kwo93}~and Drummond et al.,\cite{dru_nee09}~respectively.}
  \label{fig:corr}
\end{figure}
\begin{table}[!tb]
  \centering
  \caption{
    The correlation energy $ E_{\urm{corr}} / N $ of the 2DHEG at the selected Wigner-Seitz radii $ r_{\urm{s}} $ 
    by the FRG-DFT and the DMC method with the backflow correction.\cite{kwo93,dru_nee09}
    The number in the parentheses are errors in the last decimal place.}
  \label{tab:corr}
  \begin{tabularx}{\columnwidth}{cYYY}
    \hline
    $ r_{\urm{s}} $ ($ \mathrm{a.u.} $) & $ 1 $ & $ 5 $ & $ 10 $  \\
    \hline
    FRG-DFT & $ -0.10992 $ & $ -0.056581 $ & $ -0.039396 $ \\
    DMC\cite{kwo93} & $ -0.1096(3) $ & $ -0.0495(1) $ & $ -0.03034(2) $ \\
    DMC\cite{dru_nee09} & $ -0.1102(6) $ & $ -0.04959(3) $ & $ -0.030378(6) $ \\
    \hline
  \end{tabularx}			
\end{table}
\par
Figure \ref{fig:corr} shows the result of the $ r_{\urm{s}} $-dependence of the correlation energy $ E_{\urm{corr}} / N $ derived by the FRG-DFT. 
The computational time to derive the correlation energy from Eq.~\eqref{eq:esol} is relatively short,
which enables us to obtain the correlation energies for various $ r_{\urm{s}} $
enough to see the smooth shape of the $ r_{\urm{s}} $-dependence curve.
Concretely, we calculate the correlation energies by changing $ r_{\urm{s}} $
at intervals of $ 0.1 \, \mathrm{a.u.} $ in $ r_{\urm{s}} \geq 0.1 \, \mathrm{a.u.} $,
and at shorter intervals near $ r_{\urm{s}} = 0 $.
For comparison, Fig.~\ref{fig:corr} also shows the energies derived by the Gell-Mann-Brueckner resummation
and the Monte Carlo (MC) calculations,
which were derived from the extrapolations of the results by 
the diffusion Monte Carlo (DMC) method with the backflow correction\cite{kwo93,dru_nee09} to the infinite systems.
Since the energies given in Ref.~\onlinecite{dru_nee09} are the total energies,
the kinetic energy $ 1 / \left( 2r_{\urm{s}}^2 \right) $
and the exchange energy $ -4 \sqrt{2} / \left( 3\pi r_{\urm{s}} \right) $ are subtracted 
to extract the correlation energy.
We see that the FRG-DFT result completely reproduces the energy by the Gell-Mann-Brueckner resummation
at the small $ r_{\urm{s}} $ region, as we have discussed in Sec.~\ref{sec:exac}.
For finite $ r_{\urm{s}} $,
the FRG-DFT result seems to be relatively close to the MC results in the small $ r_{\urm{s}} $, particularly in $ r_{\urm{s}} = 1 \, \mathrm{a.u.} $,
although the deviation between the energies by the FRG-DFT and the MC simultaneously becomes larger as $ r_{\urm{s}} $ increases.
The improvement of the result by the FRG-DFT in comparison with the Gell-Mann-Brueckner resummation
at finite $ r_{\urm{s}} $ is caused by the resummation of the exchange contribution 
$ C_{\lambda = 0} \left( P \right) $ in the $ \tilde{G}^{(2)}_\lambda \left( P \right) $
performed with solving Eq.~\eqref{eq:fg2t},
since in the Gell-Mann-Brueckner resummation,
$ C_{\lambda = 0} \left( P \right) $ is not resummed but just added to the two-point density correlation function, and contributes to the energy as the second term in the right-hand side of Eq.~\eqref{eq:ecex}.
\par
Table \ref{tab:corr} shows the numerical values of the correlation energies $ E_{\urm{corr}} / N $ 
obtained from the FRG-DFT and MC calculations for several $ r_{\urm{s}} $.
At $ r_{\urm{s}} = 1 \, \mathrm{a.u.} $, we find that the FRG-DFT result agrees with both MC results within the errors of the MC calculations.
The discrepancy, however, becomes larger as $ r_{\urm{s}} $ increases:
The result of the FRG-DFT misses by approximately $ 14 \, \% $ at $ r_{\urm{s}} = 5 \, \mathrm{a.u.} $
and $ 30 \, \% $ at $ r_{\urm{s}} = 10 \, \mathrm{a.u.} $ in comparison with the MC results.
In order to improve the accuracy in large-$ r_{\urm{s}} $ region,
the inclusion of the flows of the higher-order correlation functions $ G_{\lambda}^{( n \geq 3)} $,
which are neglected in the present scheme, would be needed.
\section{Conclusion\label{sec:conc}}
\par
We have shown the first application of
the functional-renormalization-group aided density-functional theory (FRG-DFT) to the two-dimensional homogeneous electron gas.
Employing the vertex-expansion scheme up to the second order,
we have derived the correlation energy as a function of $ r_{\urm{s}} $.
We have found that the scheme reproduces the exact correlation energy at the high-density limit. 
For finite density, the resultant correlation energy 
is consistent with the results of
the Monte Carlo calculation at the high-density region,
whereas the discrepancy increases as the system becomes dilute.
\par
For more accurate description of the dilute systems, we need to improve the approximation.
An advantage of the vertex-expansion scheme is that the systematic improvement of the approximation is possible.
The next straightforward step is the inclusion of the flow of the three-point density-correlation function.
Another attractive way to take the flows of higher-order correlation functions is the KS-FRG scheme.\cite{lia18}
\par
The FRG-DFT is a flexible method and has large extensibility. 
For example, the formalism can be extended to the case when the system has arbitral spin-polarization.
The analysis of the magnetic transition in this framework is a significant future direction.
The extensions of the formalism to the three-dimensional systems
and the case of finite temperature are also straightforward.
Recently, the calculation of the density-density correlation function has been achieved in the framework of the FRG-DFT.\cite{yok18b}
Therefore, the FRG-DFT will become a tool to investigate 
not only the ground state but also excited states of the electron gas.
The superconductivity is another interesting topic regarding the electron systems, and
the inclusion of pairing fields in our framework is also an attractive future direction.
\begin{acknowledgments}
  T.~Y.~acknowledges Teiji Kunihiro and Kenichi Yoshida for their collaboration of Refs.~\onlinecite{yok18,yok18b} on which the present work is based.
  We also thank them for their interest in and valuable discussion on the present work, and useful comments on the manuscript.
  T.~Y.~was supported by the Grants-in-Aid for JSPS fellows (Grant No.~16J08574).
  T.~N.~would like to thank the RIKEN iTHEMS program, and the JSPS-NSFC Bilateral Program for Joint Research Project on Nuclear mass and life for unraveling mysteries of the r-process.
  T.~N.~also would like to thank the visitor program of the Yukawa Institute for Theoretical Physics, Kyoto University.
\end{acknowledgments}
%
%

\end{document}